\documentclass[10pt,conference,a4paper]{IEEEtran}
\usepackage[utf8]{inputenc}

\usepackage{times}

\usepackage{graphicx}
\usepackage{subfigure}
\usepackage{multirow}
\usepackage{graphicx}
\usepackage{multicol}
\usepackage{mathtools}
\usepackage{booktabs,makecell}
\DeclareGraphicsExtensions{.png,.eps,.ps,.pdf}

\usepackage{url}
\usepackage{scalerel}
\usepackage{tikz}
\usetikzlibrary{svg.path}

\definecolor{orcidlogocol}{HTML}{A6CE39}
\tikzset{
  orcidlogo/.pic={
    \fill[orcidlogocol] svg{M256,128c0,70.7-57.3,128-128,128C57.3,256,0,198.7,0,128C0,57.3,57.3,0,128,0C198.7,0,256,57.3,256,128z};
    \fill[white] svg{M86.3,186.2H70.9V79.1h15.4v48.4V186.2z}
                 svg{M108.9,79.1h41.6c39.6,0,57,28.3,57,53.6c0,27.5-21.5,53.6-56.8,53.6h-41.8V79.1z M124.3,172.4h24.5c34.9,0,42.9-26.5,42.9-39.7c0-21.5-13.7-39.7-43.7-39.7h-23.7V172.4z}
                 svg{M88.7,56.8c0,5.5-4.5,10.1-10.1,10.1c-5.6,0-10.1-4.6-10.1-10.1c0-5.6,4.5-10.1,10.1-10.1C84.2,46.7,88.7,51.3,88.7,56.8z};
  }
}

\newcommand\orcidicon[1]{\href{https://orcid.org/#1}{\mbox{\scalerel*{
\begin{tikzpicture}[yscale=-1,transform shape]
\pic{orcidlogo};
\end{tikzpicture}
}{|}}}}

\usepackage{hyperref}

\usepackage[english]{babel}

\begin{document}

%%% Título
\title{Conditional Generative Adversarial Network for keystroke presentation attack}

%%% Autores
\author{\IEEEauthorblockN{Idoia Eizaguirre-Peral\orcidicon{0000-0003-0174-6077}}
\IEEEauthorblockA{Vicomtech, Basque Research and\\Technology Alliance (BRTA)\\
20009 Donostia/San Sebastian, Spain\\
ieizagirre@vicomtech.org}
\and
\IEEEauthorblockN{Lander Segurola-Gil\orcidicon{0000-0003-4278-9081}}
\IEEEauthorblockA{Vicomtech, Basque Research and\\Technology Alliance (BRTA)\\
20009 Donostia/San Sebastian, Spain\\
lsegurola@vicomtech.org}
\and
\IEEEauthorblockN{Francesco Zola\orcidicon{0000-0002-1733-5515}}
\IEEEauthorblockA{Vicomtech, Basque Research and\\Technology Alliance (BRTA)\\
20009 Donostia/San Sebastian, Spain\\
fzola@vicomtech.org}}

\maketitle

%%% Resumen
\begin{abstract}
Cybersecurity is a crucial step in data protection to ensure user security and personal data privacy. In this sense, many companies have started to control and restrict access to their data using authentication systems. However, these traditional authentication methods, are not enough for ensuring data protection, and for this reason, behavioral biometrics have gained importance. Despite their promising results and the wide range of applications, biometric systems have shown to be vulnerable to malicious attacks, such as Presentation Attacks. For this reason, in this work, we propose to study a new approach aiming to deploy a presentation attack towards a keystroke authentication system. Our idea is to use Conditional Generative Adversarial Networks (cGAN) for generating synthetic keystroke data that can be used for impersonating an authorized user. These synthetic data are generated following two different real use cases, one in which the order of the typed words is known (ordered dynamic) and the other in which this order is unknown (no-ordered dynamic). Finally, both keystroke dynamics (ordered and no-ordered) are validated using an external keystroke authentication system. Results indicate that the cGAN can effectively generate keystroke dynamics patterns that can be used for deceiving keystroke authentication systems.

\end{abstract}

%%% Palabras clave

%%% Tipo de contribución: seleccione lo que proceda, eliminando el resto del texto
{\bf Contribution type:}  {\it Research in development } 

%%% Consejos generales 
\section{Introduction}
In data protection, cybersecurity plays a key role in authenticating users and giving them access to personal and untransferable information. A failure in the user authentication system can lead to big economic, reputational, and social damage \cite{Clarck-DA:damage}, and therefore, authentication systems are becoming more and more robust. In this scenario, biometry is playing an important role because it enables a universal, singular, permanent in time and measurable system of authenticating users \cite{Matyas-R:advantages}.

In the last years, user authentication systems based on biometry are being used in very diverse scenarios, such as in airport scanners, banking, military access control, smartphones, and forensics among others \cite{Muley-K:banking}, \cite{Bud:faceid}. These systems, usually based on Machine Learning techniques, extract feature measurements and decide whether they correspond to the characteristics of the user that is requesting access. Biometrics is divided into two subfields: physical biometrics and behavioral biometrics. However, this study is focused on behavioral biometrics.

Despite presenting optimistic results in a wide range of applications, behavioral biometrics can be attacked in different ways \cite{marcel2014handbook}. The most common attack on a user authentication system is the Presentation Attack (PA), which consists of an attack on the biometric sensor that captures the individual's measurements \cite{Ness:PAD}. Keystroke dynamics is a type of behavioral biometrics that refers to the way a user types on a keyboard taking into account the speed, the rhythm, the common mistakes and the times spent pressing and releasing each keycode. It has been demonstrated that users can be differentiated and verified depending on their typing pattern. However, it can be very difficult to imitate such keystroke patterns by another user or bot.

%Lo comentamos sí, céntrate en lo poco que hay de PAD de comportamiento (o incluso PAD de dinámicas de tecleo)

In literature, all major achievements have been made with fixed text, and, to the best of our knowledge, free-text dynamics have not been studied at a large scale yet. Moreover, specifically in Presentation Attack with this type of data has not been studied yet. For this reason, in this project, we propose a new approach that uses free-text dynamics for learning and generalizing user keystroke behavior. Furthermore, we study how to use such information to implement a keystroke presentation attack on a biometric authenticator. In particular, we propose to use conditional Generative Adversarial Networks (cGAN) for learning keystroke dynamics and generate synthetic patterns that can deceive keystroke authentication systems.

The rest of the paper is organized as follows. In Section \ref{sec:preliminaries}, concepts regarding keystroke dynamics, generative models, and related work are introduced. In Section \ref{sec:methodology}, the proposed methodology is detailed and the data used, the metrics, the experiments, and the validation process carried out in this study are presented. In Section \ref{sec:results}, results are reported and last, Section \ref{sec:conclusions}, provides conclusions and guidelines for future work.

\section{Preliminaries}\label{sec:preliminaries}
In this section, some background concepts for understanding this study are explained. In section \ref{subsec:keystroke}, keystroke dynamics are presented, and in section \ref{subsec:gan} GANs and an extension of these are presented (Conditional Generative Adversarial Networks, cGAN), and their benefits and drawbacks are explained. Last, the dataset used in this study is presented in section \ref{subsec:data}.

\subsection{Keystroke dynamics} \label{subsec:keystroke}
Keystroke dynamics are a type of behavioral biometric that measure the speed, the rhythm, the common errors, and the typing times for pressing and releasing each keycode to identify patterns of typing IDs to identify or verify a user's identity. It has been proved that each person has their typing ID due to the habit of typing certain sequences always in a similar way, such as the user's name, password, etc \cite{bergadano-GP:keystroke}. The main advantages of using keystroke dynamics in user authentication systems are that there is no need for any special hardware because a keyboard in a device is the only requirement \cite{Hassan-FH:hardware} and it enables continuous user authentication without disturbing the user experience while using such device \cite{Monrose-R:authentication}. It is thought that each user has its own typing ID and it can be very difficult to be imitated by another user or bot \cite{4631978}

In literature about keystroke dynamics, the robustness of keystroke dynamics against synthetic falsification attacks \cite{stefan-SY:forgery} and the robustness against replay attacks (an attack that collects and re-sends data to try to spoof the system) \cite{hazan-MR:replay} have been studied and it has been shown that keystroke dynamics can be used to increase user authentication reliability using different modeling techniques such as SVM and NN \cite{yu-C:problemandsol}.

Acien et al. \cite{Acien-MVF:typenet} presented a Siamese Network capable of distinguishing whether two given keystroke sequences belonged to the same user or belonged to two different users. A Siamese Network consists of a type of NN composed of several subnetworks that are identical in architecture and weights but have got different inputs whose outputs are combined in a common loss function in charge of measuring the similarity between inputs. In this work, this model will be referenced as TypeNet.

\subsection{Generative Adversarial Network} \label{subsec:gan}
Generative Adversarial Networks (GAN) are a type of generative modeling from deep learning introduced by Ian Goodfellow \cite{Goodfellow-AMX:GAN}. They make use of two Neural Networks that compete with each other during their training: the generator ($G$) and the discriminator ($D$). The idea is that, through this adversarial training, both networks improve their performances. These networks play a zero-sum or min-max game, that is, the two networks play an adversarial game, one loses when the other wins and the other way round. This min-max game may not always reach an equilibrium, such a problem is known as \textit{non-convergence} problem. Apart from the \textit{non-convergence problem}, training a GAN may also result in \textit{mode collapse} \cite{Saxena-C:challenges} and \textit{vanishing gradient} which refer to a problem in the output of the generator and a problem with the value of the gradient during training respectively.

GANs are widely used to create or generate new samples of data that are as realistic as the samples in the original dataset. Since this new framework was proposed, GANs have been used for very diverse applications, and due to the difficulty when training, many extensions of GANs have been developed. They have been used for image creation \cite{Radford-MC:unsupervised} employing Deep Convolutional Generative Adversarial Networks (DCGAN), data augmentation using GANs \cite{Zola-LEGO:bitcoin}, image-to-image translation utilizing Conditional Generative Adversarial Networks (cGAN) \cite{Isola-ZZE:conditional} and CycleGANs \cite{Zhu-PIE:cycle}, melody generation from lyrics with Conditional LSTM-GANs \cite{Yu-SC:cLSTM-GAN}. What is more, in 2016, Zhang et al. proposed the StackGAN, another extension of GANs that transforms text descriptions into realistic images.  In 2017, Karras et al. proposed a new methodology for GANs to improve training stability and speed it up through a Progressive Growing of GANs \cite{Karras-ALL:progressive}. In the last years, new GAN extensions have been also used to generate synthetic sequential data. In 2017, Esteban et al. proposed the Recurrent GAN and the Recurrent Conditional GAN to generate synthetic multi-dimensional time series \cite{Esteban-HR:timeseriesGAN}. Among these GAN structures, in this study, we implement the Conditional GAN (or cGAN).

\textbf{Conditional Generative Adversarial Networks (cGAN)} are an extension of the GANs presented by Mehdi Mirza in 2014, \cite{Mirza-MOS:cgan}. The main difference is that cGANs can generate data from a previously specified category or class. Both the generator and the discriminator in cGANs have got an extra input, the extra information that will condition the generated sample. This condition can be a class label, a numerical value, or an embedding vector. Therefore, the generator will try to generate samples from the corresponding data distribution of that class and the discriminator will learn to decide whether the given pair (sample and condition) is a matching real sample or has been artificially generated by the generator.

The cGAN designed, implemented, and validated in this study aims to generate synthetic data on keystroke dynamics starting with two inputs: a latent vector and a character embedding. An embedding is a numerical representation of some high dimensional vector or text data into a low dimensional vector. The use of embeddings in Machine Learning (ML) models is widespread due to their contribution to making model training easier and more efficient. In this study, a pre-trained character embedding known as \textit{char2vec} has been used.

\subsection{Dataset} \label{subsec:data}
This study has been carried out using a dataset provided by the Aalto University in Finland \cite{Dhakal-FKO:dataset}, in which they collected data from $168,000$ volunteers in an online study. This dataset is a large-scale dataset that contains information on the keystroke patterns of each of the volunteers. It contains keystroke times of $15$ sentences for each volunteer. For each keystroke pressed to type the sentence, the pressing and releasing times were registered for more than $136$ million of keystrokes. 

However, in this study, as the aim is to impersonate a single user and validate it with an external model, a subset of the whole dataset has been used. On the one hand, to train the external model, TypeNet, a dataset composed of data about $25$ volunteers have been used. In this study, it will be referenced as TypeNet dataset. On the other hand, to train the cGAN, a dataset of a single user has been used, which will be referenced as single-user dataset. It is important to outline that despite having the records of the pressing and releasing timestamps of the keycodes in the dataset, in line with the literature in the area, four variables were extracted to build the ML models: Hold Latency (HL), Press Latency (PL), Release Latency (RL) and  Inter-Key Latency (IL).

\section{Methodology}\label{sec:methodology}
In this section, first, a global view of the methodology is explained, next, in section \ref{subsec:model}, the architecture of the cGAN is detailed. In section \ref{subsec:framework}, some important aspects of the experimental framework are presented. In section \ref{subsec:metrics}, the metrics used to validate the results are presented, and finally, in section \ref{subsec:validation}, the validation procedure is described.

\subsection{Proposal}\label{subsec:proposal}

This study aims to generate synthetic data on keystroke dynamics to impersonate a user and develop a presentation attack. To that end, we propose a cGAN able to learn the keystroke dynamics of a specific user and generate realistic data samples. These generated data samples will be able to fool an external model, TypeNet, used as a biometric authenticator to access a determined service.

\begin{figure}
     \centering
    \includegraphics[width=0.7\linewidth]{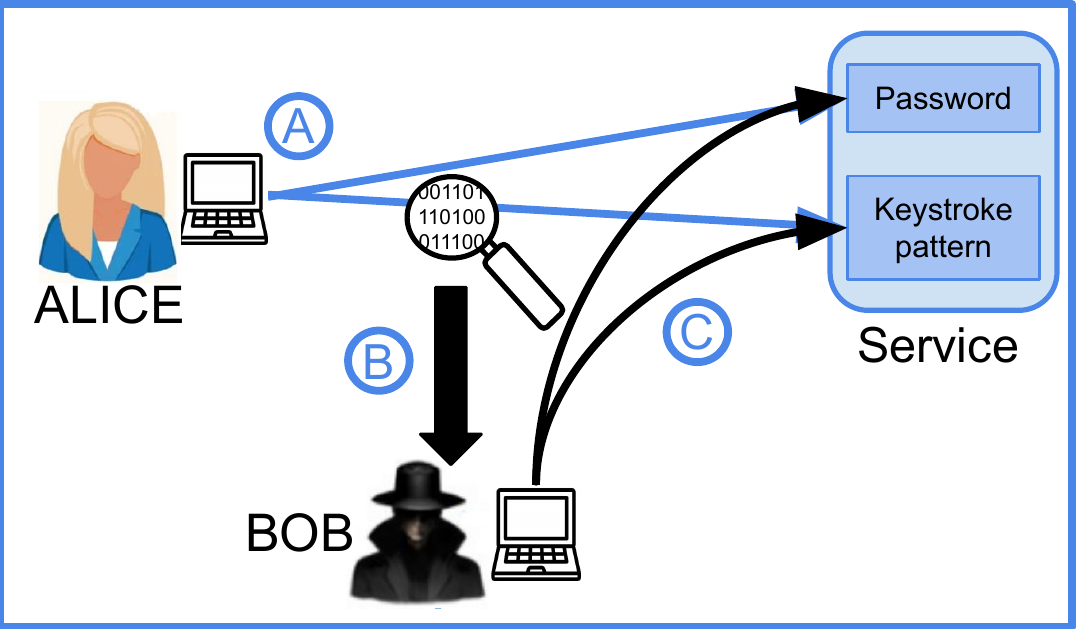}
    \caption{Steps of our Presentation Attack.}
    \label{fig:stepa}
\end{figure}

Let us assume that Alice is a registered user of concrete service. To access this service, a double verification is carried out. the user that wants to access the service, not only has to type its password, but the system also checks the keystroke dynamics of the user typing it and verifies the user's typing ID to enable access or deny, see step $A$ in Fig. \ref{fig:stepa}. However, a non-registered user, Bob, wants to access the service with Alice's credentials. To do so, let us assume Bob already knows Alice's password, see step $B$ in Fig. \ref{fig:stepa}. To access the service, he needs to imitate Alice's typing behavior. In other words, he needs to impersonate her by generating synthetic behavioral biometric data to fool the biometric authenticator, see step $C$ in Fig. \ref{fig:stepa}.

%\begin{figure}
 %    \centering
  %  \includegraphics[width=0.7\linewidth]{StepB.pdf}
   % \caption{Step B.}
    %\label{fig:stepb}
%\end{figure}
%\begin{figure}
 %    \centering
  %  \includegraphics[width=0.7\linewidth]{StepC.pdf}
   % \caption{Step C.}
    %\label{fig:stepc}
%\end{figure}

In this context, we propose a new methodology to deploy a keystroke PA using cGAN technology. Our methodology is composed of three phases. In phase $1$, the aim is to train the biometric authenticator following the process indicated in \cite{Acien-MVF:typenet}, which will be in charge of deciding whether two given $15$-character input sequences belong to a single user or belong to two different users. In phase $2$, making use of the single-user dataset and considering training data by words, the cGAN has been trained to be capable of generating realistic behavioral data on Alice's keystroke dynamics. Last, in phase $3$, the PA is carried out, hence, the synthetic data is reconstructed into synthetic sequences of several words concatenated together and are introduced in the initial biometric authenticator to impersonate Alice. This last phase has been developed in two different real use cases, one in which the order of the words in the reconstructed sequence was identical to the order in the original dataset, and the other in which words were concatenated randomly considering a space between two words.

\subsection{The model}\label{subsec:model}
The cGAN we propose is called Vanilla-cGAN because the internal architecture of the generator and the discriminator in the model are NN. The generator generates synthetic data based on a latent vector with $500$ random values that follow a Gaussian distribution and a $100$ dimensional character embedding that encodes the word whose typing times are generated. The discriminator distinguishes whether a given data sample is real or synthetic based on the typing times and the keycode values ($15\times5$) for a given word and its corresponding $100$ dimensional character embedding. 

\subsection{Experimental framework} \label{subsec:framework}
As it has been mentioned previously, data on keystroke dynamics is usually treated using $5$ variables that describe the typing behaviour of the users: \textit{HL}, \textit{PL}, \textit{RL}, \textit{IL} and the \textit{KEYCODE} in ASCII code. In this study, both models, TypeNet and the cGAN, use these variables as training data. The difference lies in the sequences that each model receives in the training, and consequently, in the validation process. Note that the biometric authenticator considers sequences of $15$ characters whereas the cGAN considers the training by words stored in sequences with the same length but with extra padding when the word has less than $15$ characters.

The training process is carried out until reaching concrete stopping criteria related to the precision of the GAN discriminator. In particular, every $50$ epoch, the training has been paused and $5$ subsets of $32$ real and $32$ synthetic samples have been used as input for the discriminator to measure its performance in classifying the given samples. If the accuracy of the discriminator has been above $85$\% for both real and synthetic samples, the training process is stopped, otherwise, it is resumed until the next validation (on the next $50$ epochs).

Once the training of the cGAN is completed, in phase $3$, to perform a presentation attack, two use cases or conditions have been considered: 
\begin{itemize}
    \item \underline{Condition $1$:} in which the order of the words in the reconstructed sequence corresponds to the same order in the single-user dataset; 
    \item \underline{Conditions $2$:} in which the order of the words in the reconstructed sequence is done randomly considering a space between every other word in the sequence.
\end{itemize}

\subsection{Metrics} \label{subsec:metrics}
Let us consider the confusion matrix for a binary classification composed of the following values: True Negatives (TN), False Negatives, (FN) False Positives (FP), and True Positives (TP).

In this study the evaluation metrics used to determine the quality of the obtained results are the following:
\begin{itemize}
    \item \underline{Accuracy}: 
        \begin{equation}
        Acc=\frac{TP+TN}{TP+TN+FP+FN}
        \end{equation}
    \item \underline{Recall}: 
        \begin{equation}
        Rec=\frac{TP}{FN+TP}
        \end{equation}
    \item \underline{Precision}:
        \begin{equation}
        Pre=\frac{TP}{TP+FP}
        \end{equation}
    \item \underline{F1-Score}: 
        \begin{equation}
        F1=\frac{2\times TP}{2\times TP+FN+FP}
        \end{equation}
    \item \underline{Matthews Correlation Coefficient (MCC)}: 
        \begin{equation*}
            \centering
            MCC=
        \end{equation*}
        \begin{equation}
            \resizebox{.9\hsize}{!}{$\frac{TN\times TP-FP \times FN}{\sqrt{(TN+FN)\times(FP+TP)\times(TN+FP)\times(FN+TP)}}$}
        \end{equation}
\end{itemize}

\subsection{Validation} \label{subsec:validation}
As described in phase $3$ of our methodology in section \ref{sec:methodology}, the validation has been done using the biometric authenticator, TypeNet, employing three tests:
\begin{itemize}
    \item \underline{Test $1$: Real vs. Fake} (\textit{Acc\_rf}):  $20$ real sequences and $20$ generated sequences from the \underline{same user} are validated to try to fool the external model.
    \item \underline{Test $2$: Fake vs. Fake} (\textit{Acc\_ff}):  $20$ generated sequences used in test $1$ and another $20$ generated sequences from the \underline{same user} are validated with the external model.
     \underline{Test $3$: Real other vs. Fake} (\textit{Acc\_rof}). $20$ generated sequences used in test $1$ and $20$ real sequences that belong to other users (the other $24$ users for which TypeNet has been trained) are validated with the external model.
\end{itemize}
\section{Results} \label{sec:results}
Tab \ref{tab:res1} show the accuracy values of the three tests in both use cases. 
It can be observed that in general, the values are high for all tests. Note that as the values in tests $1$ and $2$ are similar, it can be concluded that the synthetic samples are realistic, and the biometric authentication is not able to detect differences at all. With the results in test $3$, it can be concluded that the generated samples for Alice are not similar to any other user in the TypeNet dataset. In what differences between conditions concern, it can be concluded that the order of the words does not affect at all the typing behavior of the users.
\begin{table}[htb]
    \centering
    \resizebox{\linewidth}{!}{%
    \begin{tabular}{c|ccc}
    \hline
    \hline
                        & \textbf{\begin{tabular}[c]{@{}c@{}}Test 1:\\ Real vs. Fake\end{tabular}} & \textbf{\begin{tabular}[c]{@{}c@{}}Test 2:\\ Fake vs. Fake\end{tabular}} & \textbf{\begin{tabular}[c]{@{}c@{}}Test 3:\\ Real other \\ vs. Fake\end{tabular}} \\ \hline
    \textbf{Condition 1} & 0.945                                                                    & 0.945                                                                    & 0.975                                                                             \\ \hline
    \textbf{Condition 2} & 0.950                                                                    & 0.960                                                                    & 0.955                                                                             \\ \hline
    \hline
    \end{tabular}%
    }
    \caption{Accuracy of the classification of TypeNet for the generated data.}
    \label{tab:res1}
\end{table}

\section{Conclusions} \label{sec:conclusions}
The main goal of this work has been to generate synthetic data on keystroke dynamics in an adversarial way and to deploy a Presentation Attack. To validate the results, the performance of the discriminator has been measured and an external model has been used. Results indicate that keystroke dynamics patterns can be adversarially generated and promising results can be obtained in the two different real scenarios. This project shows the relevance of typing behavior data generation using adversarial networks with several approaches. It has been proved that valid data to impersonate users can be generated using generative modeling from ML. 

Further research is needed to determine whether there is any other GAN extension that would reduce training time and improve results' performance.

%%% inclusión de referencias
\bibliographystyle{IEEEtran}

%\externalbibliography{yes}
%\bibliography{your_external_BibTeX_file}
\bibliography{main.bib}

\end{document}